\documentclass[epj,final]{svjour} 

\usepackage{graphics}
\usepackage{epsf}
\usepackage{epsfig}

\begin{document} 

\flushbottom
\def\bottomfraction{0.5}

\title{Persistence distributions in a conserved lattice gas with absorbing states}

\author{S. L\"ubeck \and A. Misra}

\institute{
Theoretische Tieftemperaturphysik, 
Gerhard-Mercator-Universit\"at Duisburg,\\ 
Lotharstr. 1, 47048 Duisburg, Germany \\}

\date{received 27 August 2001}

\abstract{
Extensive simulations are performed to study the persistence behavior
of a conserved lattice gas model 
exhibiting an absorbing phase transition from an
active phase into an inactive phase.
Both the global and the local persistence exponents
are determined in two and higher dimensions. 
The local persistence exponent obeys a scaling relation 
involving the order parameter exponent of the absorbing phase transition. 
Furthermore we observe that the global persistence exponent exceeds 
its local counterpart in all dimensions in contrast
to the known persistence behavior in reversible
phase transitions.
}


\maketitle

\section{Introduction}

In the last decade numerous articles have been
devoted to the study of persistence behavior in nonequilibrium 
systems~\cite{DERRIDA_1,BRAY_1,STAUFFER_1,MAJUMDAR_4,CUEILLE_1,MAJUMDAR_5,DERRIDA_2,LEE_1,OERDING_1,HINRICHSEN_2,ALBANO_1}.
The basic quantity of interest is the persistence 
distribution~$P(t)$, defined as the probability that a
certain physical quantity does not
change its state up to time~$t$ during a stochastic
evolution.
The persistence distribution
has been observed to decay as $P(t) \sim t^{-\theta }$,
where $\theta$ is denoted as the persistence exponent. 
Since the persistence behavior depends on the entire history 
of evolution of the nonequilibrium process, there are only
few instances of an exact calculation of the persistence 
exponent in the literature~\cite{BRAY_1,DERRIDA_2,LEE_1}. 
In general the persistence exponent is nontrivial in the sense
that it cannot be obtained from scaling relations,
except of certain Markovian systems~\cite{MAJUMDAR_4}.
There exists two different types of persistence 
exponent $\theta_{\rm l}$ and $\theta_{\rm g}$ associated with 
the persistence behavior of the local and the global order 
parameter, respectively. 
These two exponents have been observed in general to 
be different from each other. 
Usually for reversible phase transitions the 
inequality $\theta_{\rm g}<\theta_{\rm l}$ is observed, however in the case
of the continuous absorbing phase transition of the 
Ziff-Gulari-Barshad model~\cite{ZIFF_1},
which belongs to the universality class of directed percolation, 
the reverse inequality has been observed numerically~\cite{ALBANO_1}.

Recently Rossi {\it et al.}~introduced a conserved
lattice gas (CLG) model that exhibits a continuous phase transition
to an absorbing state at a critical value of the
particle density~\cite{ROSSI_1}.
Numerical simulations in various dimensions~\cite{LUEB_19}
confirm the conjecture of~\cite{ROSSI_1} that the CLG model
belongs to a new universality class different from the
well known universality class of directed percolation 
(see ~\cite{HINRICHSEN_1} for an overview).
In this work we consider 
the persistence distributions of the CLG model
in various dimensions ($D=2,3,4,5$).
We determine numerically the global and local
persistence exponents as well as the exponent
$\nu_{\scriptscriptstyle \parallel}$ which describes
the divergence of the correlation time
at the transition point.
To our knowledge this is the first time that the 
persistence behavior of a non-equilibrium critical system
with absorbing states is investigated in higher dimensions, 
i.e., below and above the upper critical dimension~$D_{\rm c}$.
Similar to~\cite{ALBANO_1} we observe the reverse
inequality $\theta_{\rm g} > \theta_{\rm l}$
for the CLG model. 
We believe that systems exhibiting irreversible phase 
transitions into an absorbing state
are generally characterized by this
reversed inequality in contrast to systems exhibiting 
a reversible phase transition, in particular 
equilibrium phase transitions.

The paper is organized as follows. 
In the next section we describe the CLG model. 
In section~\ref{sec:pers_dist} we define the local and the global
persistence distributions for this model and describe their 
steady state scaling behavior in the critical regime.
In the following two sections we present the 
determination of the global (sec.~\ref{sec:global})
and local (sec.~\ref{sec:local}) persistence exponents.
With a short discussion we conclude the paper in 
section~\ref{sec:dis}.

\section{The CLG model}
\label{sec:model}

We consider the CLG model (see~\cite{ROSSI_1}) 
on $D$-dimensional cubic lattices of linear size~$L$.
Initially one distributes randomly $N=\rho L$ 
particles on the system where $\rho$ denotes the
particle density.
In order to mimic a repulsive interaction
a given particle is considered as {\it active} if at least one of its 
$2D$ neighboring sites on the cubic lattice is occupied
by another particle.
If all neighboring sites are empty the particle
remains {\it inactive}.
Active particles are moved in the next update step to one 
of their empty nearest neighbor sites, selected at random.
Starting from a random distribution of particles
the system reaches after a transient regime a
steady state which is characterized by the
density of active sites $\rho_{\rm a}$.
The average density $\langle \rho_{\rm a} \rangle$ is the order 
parameter of the absorbing phase transition, i.e., it vanishes 
if the control parameter $\rho$ is lower
than the critical value $\rho_{\rm c}$.
Approaching the critical value from above the 
order parameter scales as
$\langle \rho_{\rm a} \rangle \sim \delta \rho^\beta$,
where $\delta \rho=\rho - \rho_{\rm c}$.
The order parameter exponent~$\beta$ as well as the
scaling behavior of the order parameter fluctuations
are investigated numerically in various dimensions
in~\cite{LUEB_19}.
In our simulations, 
a random sequential update scheme is applied, i.e.,
all active sites are updated at each time step 
in a randomly chosen sequence.
The considered system sizes are
$L\le 1024$ for $D=2$, $L\le 128$ ($D=3$),
$L\le 32$ ($D=4$) and $L\le 16$ in the case of $D=5$. 
All measurements are performed in the steady state and the
used equilibration procedures are discussed in detail
in~\cite{LUEB_19}.

\section{Persistence distributions}
\label{sec:pers_dist}

In this work we consider the persistence behavior of the CLG model.
Usually one distinguishes the so-called global 
persistence distribution $P_{\rm g}(t)$ and
the local persistence distributions $P_{\rm l}(t)$.
The global persistence characterizes the time evolution
of the order parameter, i.e., it characterizes how long the
whole system remains in a given phase.
For instance in magnetic systems $P_{\rm g}(t)$ can be defined
as the probability that the magnetization changes
sign after the duration $t$. 
At criticality $P_{\rm g}(t)$ is expected 
(see for instance~\cite{MAJUMDAR_5})
to decay algebraically 
\begin{equation}
P_{\rm g}(t) \; \sim \; t^{-\theta_{\rm g}},
\label{eq:pers_global_01}
\end{equation}
where the time $t$ corresponds to the number of update steps.
Off criticality, the finite correlation time 
$\xi_{\scriptscriptstyle \parallel}$
limits this power-law behavior and the corresponding
probability distribution displays 
a cut-off at $t = \cal{O}(\xi_{\scriptscriptstyle \parallel})$.
Using 
$\xi_{\scriptscriptstyle \parallel} \sim 
\delta\rho^{-\nu_{\scriptscriptstyle \parallel}}$ 
and standard scaling arguments one expects
that the persistence probability obeys the ansatz
\begin{equation}
P_{\rm g}(t) \; = \;
\delta \rho^{\nu_\parallel 
\theta_{\rm g}} \; {\cal P}_{\rm g}
(\delta \rho^{\nu_\parallel} \, t),
\label{eq:pers_global_02}
\end{equation}
with the universal function~${\cal P}_{\rm g}$.
Thus the scaling behavior of the probability 
distribution $P_{\rm g}(t)$
allows to determine beside the persistence exponent~$\theta_{\rm g}$ 
the exponent~$\nu_{\scriptscriptstyle \parallel}$ of the temporal 
correlations in all considered dimensions.

The phase transition of the CLG model is an irreversible transition
into an absorbing state, i.e., the order parameter 
$\langle \rho_{\rm a} \rangle$
is a strictly positive quantity.
In this cases it is usual~\cite{HINRICHSEN_2,ALBANO_1} to 
measure the probability
that the deviation of the density of active
sites from the order parameter, 
i.e.~$\rho_{\rm a}-\langle \rho_{\rm a} \rangle$, changes sign 
after $t$ update steps.
Furthermore, the distribution of the deviations could be 
different below and above $\langle \rho_{\rm a} \rangle$.
In order to take this asymmetry into account we
calculate two different quantities 
$P_{\rm g}^{\scriptscriptstyle >}(t)$ and
$P_{\rm g}^{\scriptscriptstyle <}(t)$ corresponding to the 
global persistence distribution for 
$\rho_{\rm a}(t)> \langle \rho_{\rm a} \rangle$ and
$\rho_{\rm a}(t)< \langle \rho_{\rm a} \rangle$, respectively.

The so-called local persistence is defined as the
probability that the local order parameter at a given site,
for instance a single spin, remains in a phase, i.e.,
it does not flip within a duration $t$.
In the case of the CLG model we consider
the steady state probability distribution~$P_{\rm l}(t)$ that a 
particle remains inactive for exactly $t$ update
steps.
Since no spontaneous generation of active sites
is allowed in the CLG model, a given site may remain 
inactive for a long time until it is re-activated in the
presence of another active site.
Thus one expects that the corresponding probability 
distribution decays algebraically with an 
exponent $\theta_{\rm l}$,
\begin{equation}
P_{\rm l}(t) \; \sim \; t^{-\theta_{\rm l}}.
\label{eq:pers_local_01}
\end{equation}
Alternatively one may ask how long a given particle remains active.
But active sites can be spontaneously "annihilated" to
become inactive and the corresponding probability 
distribution decays exponentially.
This "asymmetry"~\cite{HINRICHSEN_3} in the dynamical 
behavior of active and 
inactive sites is already known from other systems
exhibiting an absorbing phase transition, for instance
directed percolation.

Above the transition point ($\rho > \rho_{\rm c}$) the 
local persistence distribution displays a cut-off
at $t={\cal O}(\xi_{\rm l})$, where
$\xi_{\rm l}$ denotes the correlation time of the effective 
single site process.
Assuming that $\xi_{\rm l}$ scales as 
$\xi_{\rm l} \sim \delta\rho^{-\nu_{\rm l}}$ one
obtains the scaling behavior of the local
persistence distribution
\begin{equation}
P_{\rm l}(t) \; = \;
\delta \rho^{\nu_{\rm l} 
\theta_{\rm l}} \; {\cal P}_{\rm l}
(\delta \rho^{\nu_{\rm l}} \, t)
\label{eq:pers_local_02}
\end{equation}
with the local critical exponents $\theta_{\rm l}$ and 
$\nu_{\rm l}$.
These local exponents are connected to the order
parameter exponent~$\beta$.
Using the scaling ansatz Eq.\,(\ref{eq:pers_local_02})
one can calculate the average time between two
activation events
\begin{equation}
\langle t \rangle  \; = \; 
\int \, t \, P_{\rm l}(t) \, {\rm d}t
\;  \sim \;  \delta\rho^{-\nu_{\rm l}(2-\theta_{\rm l})}.
\label{eq:pers_local_average}
\end{equation}
Approaching the transition point $(\delta \rho \to 0)$
the average time between two activation events diverges
(we will see that $\theta_{\rm l}<2$ in all dimensions).
The number of activation events within a certain duration
is proportional to ~$1/\langle t \rangle$.
On the other hand the number of 
activations is proportional to the probability that a
site is active at a given time. Therefore
$1/\langle t \rangle \propto \langle \rho_{\rm a} \rangle $
and this leads to the scaling relation
\begin{equation}
\nu_{\rm l}\, (2-\theta_{\rm l}) \; = \; \beta.
\label{eq:pers_local_scal_rel}
\end{equation}

In spatially extended systems the time evolution of 
a single site is coupled to that of its neighbors,
thus the effective single site evolution is 
non-Markovian.
In this sense the local persistence distribution
is related to an infinite point correlation 
function (see for instance~\cite{ALBANO_1,HINRICHSEN_3})
whereas the persistence distribution of the order parameter 
is related to a two-point correlation function.
Thus one expects that the exponent $\nu_{\rm l}$
does not agree with the exponent of the order parameter
$\nu_{\rm l} \neq \nu_{\scriptscriptstyle \parallel}$.
But this inequality does not imply that the order 
parameter scaling behavior of the CLG model is
characterized by an additional second correlation
time because $\nu_{\rm l}$ just describes the temporal
correlations of an effective single site time-series
and not of the temporal evolution of the order parameter.
Furthermore, one may expect for the same reason that 
the persistence exponents are different, 
i.e.,  $\theta_{\rm g} \neq \theta_{\rm l}$.
Previous investigations suggest a general behavior
with $\theta_{\rm g} < \theta_{\rm l}$.
In this case the persistence behavior
of the global properties are stronger than those of the
local properties.
For instance the average duration between two spin flips
in ferromagnetic systems is smaller than the 
duration between order parameter flips.
Recently, a violation of this inequality was reported
by Hinrichsen and Koduvely who observed 
$\theta_{\rm g} \approx \theta_{\rm l}$ in the case
of directed percolation~\cite{HINRICHSEN_3}.
Additionally the persistence distributions of the
Ziff-Gulari-Barshad model~\cite{ZIFF_1}, 
which exhibits a continuous absorbing phase transition
belonging to the directed percolation universality class, 
are characterized in the "long-time regime"~\cite{ALBANO_1}
by the inequality $\theta_{\rm g} > \theta_{\rm l}$ for $D=2$.

\section{Global persistence distributions}
\label{sec:global}

\begin{figure}[b]
 \epsfxsize=8.6cm
 \epsffile{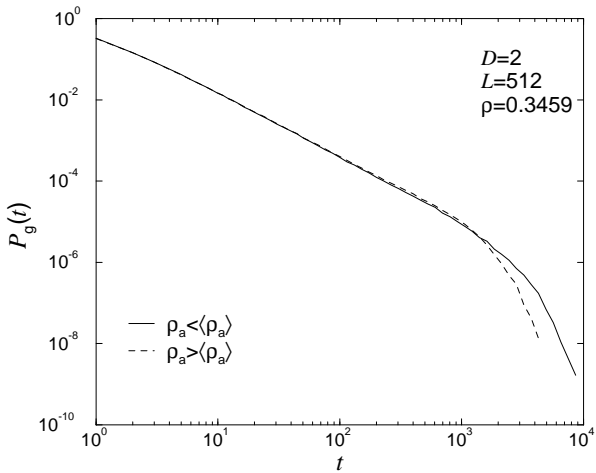}
  \caption{
    The global persistence distributions $P_{\rm g}^{\scriptscriptstyle <}(t)$
    and $P_{\rm g}^{\scriptscriptstyle >}(t)$ for a given system size and 
    particle density.  
    The different asymmetric behavior below and above the average
    value $\langle \rho_{\rm a} \rangle$ (see text) can be clearly
    seen in the cut-off behavior.
   }
 \label{fig:glo_2d} 
\end{figure}

We consider the global persistence distributions 
for dimensions $D=2,3,4$.
As explained above one expects a cut-off of the
probability distributions for 
$t\approx {\cal O} (\xi_{\scriptscriptstyle \parallel})$.
Thus, ${\cal O}(\xi_{\scriptscriptstyle \parallel})$ update
steps have to be performed to measure one of these 
long-time events.
This shows that extensive simulations are required
to obtain the global probability distribution
with a sufficient accuracy.
For instance $5\times10^6$ single update steps were
used in~\cite{LUEB_19} to determine the critical
behavior of the order 
parameter~$\langle \rho_{\rm a} \rangle$ for $D=2$.
Here, we used for similar system parameters 
$4\times10^7$ update steps in the steady state which yield approximately 
$6\times10^5$ single 
events for the measurement of the global probability
distribution.

\begin{figure}[t]
 \epsfxsize=8.6cm
 \epsffile{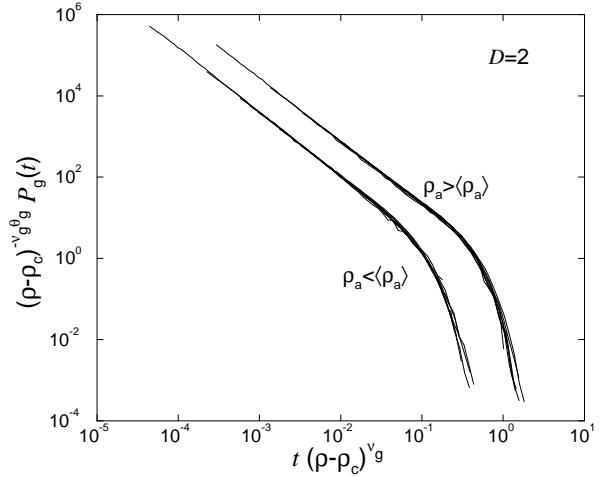}
  \caption{
    Scaling analysis of the two-dimensional global
    persistence distributions $P_{\rm g}^{\scriptscriptstyle <}(t)$
    and $P_{\rm g}^{\scriptscriptstyle >}(t)$.
    In the case of $P_{\rm g}^{\scriptscriptstyle <}(t)$ the curves are 
    shifted in the lower-left direction.
    The values of the exponents are presented in the text.
   }
 \label{fig:pers_global_02} 
\end{figure}

Fig.\,\ref{fig:glo_2d} shows the distributions
$P_{\rm g}^{\scriptscriptstyle >}(t)$ 
and $P_{\rm g}^{\scriptscriptstyle <}(t)$
for the two-dimensional model in a log-log plot.
As one can see the slopes of the curves are nearly
the same but the cuf-off behavior differs significantly.
Simulations for different values of $\delta \rho$
are performed and the obtained curves are analyzed 
according to the scaling ansatz Eq.\,(\ref{eq:pers_global_02}).
In the case of the two-dimensional model we
obtain a good data collapse for 
$\theta_{\rm g}^{\scriptscriptstyle >}=1.57\pm 0.05$, 
$\nu_{\scriptscriptstyle \parallel}^{\scriptscriptstyle >}=1.15\pm0.04$
and
$\theta_{\rm g}^{\scriptscriptstyle <}=1.60\pm 0.06$, 
$\nu_{\scriptscriptstyle \parallel}^{\scriptscriptstyle <}=1.18\pm0.05$,
respectively.
The corresponding curves are shown in Fig.\,\ref{fig:pers_global_02}.
Although the distributions $P_{\rm g}^{\scriptscriptstyle >}(t)$ 
and $P_{\rm g}^{\scriptscriptstyle <}(t)$
differ at the cut-offs the scaling behavior, i.e.~the exponents,
of both distributions agree within the error-bars.

\begin{figure}[t]
 \epsfxsize=8.6cm
 \epsffile{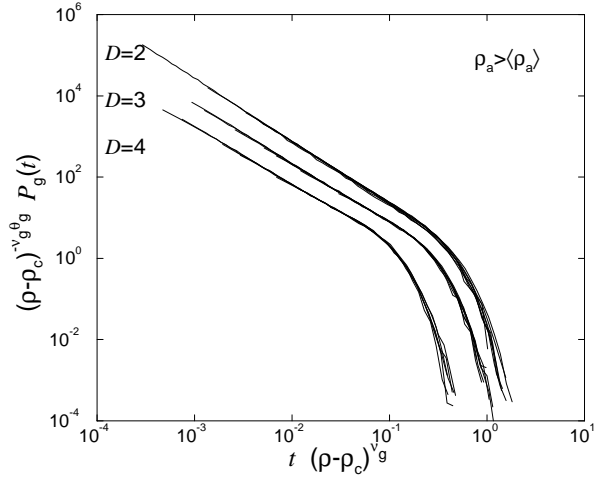}
  \caption{
    Scaling analysis of the global persistence 
    distribution $P_{\rm g}^{\scriptscriptstyle >}(t)$ for various 
    dimensions.
    With increasing dimensions the slope of the
    curves (i.e.~the exponent $\theta_{\rm g}$)
    decreases slightly.
    The curves of the three- and four-dimensional
    system are shifted in the lower-left
    direction.
    The corresponding data-collapses for 
    $P_{\rm g}^{\scriptscriptstyle <}(t)$ are of
    similar accuracy (not shown).
   }
 \label{fig:pers_global_01} 
\end{figure}

Our simulations yield that the different 
cut-off behavior of $P_{\rm g}^{\scriptscriptstyle >}(t)$ 
and $P_{\rm g}^{\scriptscriptstyle <}(t)$
vanishes with increasing dimension.
Therefore, no significant differences between the exponents
$\theta_{\rm g}^{\scriptscriptstyle >,<}$ and 
$\nu_{\scriptscriptstyle \parallel}^{\scriptscriptstyle >,<}$
are observed in higher dimensions too.
In the case of the three dimensional model 
sufficient data collapses (see Fig.~\ref{fig:pers_global_01})
are obtained for the values
$\theta_{\rm g}^{\scriptscriptstyle >}=1.53\pm 0.06$, 
$\nu_{\scriptscriptstyle \parallel}^{\scriptscriptstyle >}=1.03\pm0.03$
and
$\theta_{\rm g}^{\scriptscriptstyle<}=1.54\pm 0.06$, 
$\nu_{\scriptscriptstyle \parallel}^{\scriptscriptstyle <}=1.05\pm0.03$.

It is known that that the upper critical dimension
of the CLG is $D_{\rm c}=4$~\cite{LUEB_19}.
Thus one expects that a data-collapse of the corresponding
curves is obtained if one uses the mean-field values
of the exponent, i.e., $\nu_{\scriptscriptstyle \parallel}=1$
and $\theta_{\rm g}=3/2$.
In the latter case we assume that the time series of the 
order parameter can be mapped to a generalized random walk.
Using these values a good data-collapse is obtained 
which is plotted in Fig.\,\ref{fig:pers_global_01}.

\section{Local persistence distributions}
\label{sec:local}

\begin{figure}[t]
 \epsfxsize=8.6cm
 \epsffile{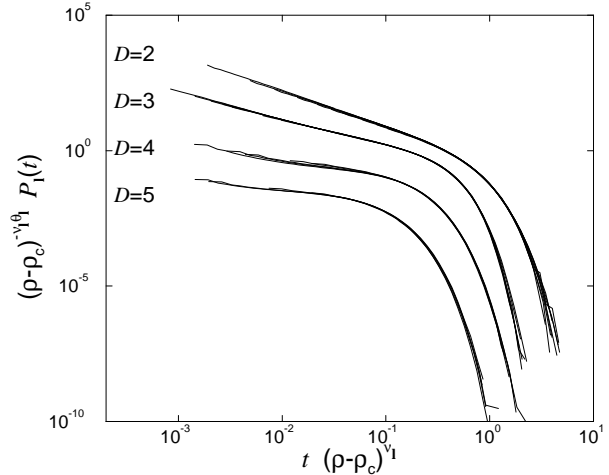}
  \caption{
    Scaling analysis of the local persistence 
    distribution $P_{\rm l}(t)$ for various 
    dimensions.
    With increasing dimensions the slope of the
    curves (i.e.~the exponent $\theta_{\rm l}$)
    decreases. 
    For $D=3,4,5$ the corresponding curves are 
    shifted in the lower-left direction.
   }
 \label{fig:pers_local_01} 
\end{figure}

We consider
the local persistence distribution $P_{\rm l}(t)$ 
of the CLG model for $D=2,3,4,5$.
Compared to the global persistence fewer
update steps are needed to determine the local persistence
distribution for a similar degree of accuracy because
the time evolution of all particles ($N=L^D \rho$) 
can be used to measure $P_{\rm l}(t)$.
Similar to the above analysis one varies again
the exponents $\theta_{\rm l}$ and $\nu_{\rm l}$
until one gets a data-collapse of different curves
which correspond to different values of $\delta \rho$.
The resulting data-collapses for $D=2,3,4,5$ are
plotted in Fig.\,\ref{fig:pers_local_01}. 
The local persistence exponent decreases from 
$\theta_{\rm l}=1.41\pm 0.02$ for $D=2$ to 
$\theta_{\rm l}=1.19\pm 0.03$ for $D=3$.
For $D\ge 4$ good data-collapses 
are obtained for $\theta_{\rm l}=1$, i.e., 
our results suggest that 
the mean-field value is $\theta_{\rm l}=1$.

The values of the local correlation exponent $\nu_{\rm l}$
decreases from $\nu_{\rm l}=1.10\pm 0.02$ ($D=2$) to
$\nu_{\rm l}=1.04\pm 0.03$ ($D=3$).
In the case of $D=4$ we used $\nu_{\rm l}=1$ to obtain
the data-collapse.
Surprisingly a significantly different value $\nu_{\rm l}=1.08\pm 0.03$
has to be used to produce a data-collapse for $D=5$ 
(see Fig.\,\ref{fig:pers_local_01}).
But in the case of the five-dimensional system the
simulations are limited by the available 
system sizes~($L\leq 16$).
Therefore, it is possible that the above value describes
a certain transient regime instead of the real critical
scaling behavior.
Therefore, we think that $\nu_{\rm l}=1$ describes the
actual mean-field behavior.

In the following we check the scaling 
relation~Eq.\,(\ref{eq:pers_local_scal_rel})
which connects the exponents $\beta$, $\nu_{\rm l}$ and
$\theta_{\rm l}$ in the steady state.
For $D=2$ our numerically obtained result 
$\nu_{\rm l}(2-\theta_{\rm l})=0.649$ agrees with the 
result $\beta=0.637\pm 0.009$ from~\cite{LUEB_19}.
In the case of $D=3$ we get 
$\nu_{\rm l}(2-\theta_{\rm l})=0.842$ 
which agrees again with $\beta=0.837\pm 0.015$ obtained 
from the same reference.
Finally the values $\nu_{\rm l}=1$, $\theta_{\rm l}=1$,
and $\beta=1$~\cite{LUEB_19,LUEB_20} fulfill the
scaling relation exactly which confirms the above conclusion
that $\nu_{\rm l}=1$ is valid for $D\ge 4$.

It is worth considering the average number of activations
within a certain time 
($n_{\rm act} \sim 1/\langle t \rangle$) in detail.
Using the 
Eqs.\,(\ref{eq:pers_local_average},\ref{eq:pers_local_scal_rel})
one sees that $n_{\rm act}$ decreases with the
dimension, i.e.,
$n_{\rm act}^{\scriptscriptstyle (D)} > n_{\rm act}^{\scriptscriptstyle (D+1)}$
for $\delta\rho \to 0$ because
$\beta^{\scriptscriptstyle (D)} < \beta^{\scriptscriptstyle (D+1)}$.
This suggests that the probability to re-activate
a given site decreases with increasing dimension.
This result could be surprising at first sight
since one would expect that due to the increasing
number of next neighbors in higher dimensions
it is more likely to re-activate a given lattice site.
But the increasing number of next neighbors
results in a reduced re-activation probability.
We guess that this behavior is caused by a
reduced return probability of activations,
similar to a random walk.
The probability that a symmetric random walk
ever returns to its origin is certain for
$D=1$ and $D=2$ but uncertain for $D=3$ (see
for instance~\cite{GRIMMETT_1}).
In the case of the CLG model one can assume
that close to criticality only
few activations diffuse through the system.
Increasing the dimension it becomes less likely
that these activations return to their origins.

\section{Discussion}
\label{sec:dis}

Considering the results of the local and global
persistence exponents (see Table\,\ref{table:exponents})
one concludes that the CLG model
is characterized by the reversed persistence
behavior $\theta_{\rm g} > \theta_{\rm l}$.
The usual behavior 
$\theta_{\rm g}< \theta_{\rm l}$ was
observed in reversible equilibrium phase transitions
and means that the global order parameter persists longer in
a given phase than the corresponding local quantities.
For instance it is more likely in a magnetic system
that a single spin flips than the magnetization.

Our result and the similar one of~\cite{ALBANO_1} 
suggest that critical systems with absorbing states behaves
generally different from reversible equilibrium systems.
Here, the order parameter fluctuates on smaller scales
than the local sites.
Approaching the transition point ($\delta \rho \to 0$) 
only a few sites are activated.
Considering snapshots of the CLG model we observed
that these activated sites are concentrated
around small regions of activation which "diffuses" very
slowly through the system.
Thus non-active particles persist for a long time
in their state until they are re-activated by other active particles.
But the order parameter fluctuates around its mean
value within this time.
This behavior is a common feature of absorbing
phase transitions and we therefore think that 
systems exhibiting an absorbing phase transitions are 
generally characterized by the reversed persistence 
behavior
$\theta_{\rm g} > \theta_{\rm l}$.



\begin{thebibliography}{10}

\bibitem{DERRIDA_1}
B. Derrida, {A.\,J.~Bray}, and {C.~Godr\`{e}che}, J.~Phys.~A {\bf 27},  L357
  (1994).

\bibitem{BRAY_1}
{A.\,J.~Bray}, B. Derrida, and {C.~Godr\`{e}che}, Europhys.~Lett. {\bf 27},
  175  (1994).

\bibitem{STAUFFER_1}
D. Stauffer, J.~Phys.~A {\bf 27},  5029  (1994).

\bibitem{MAJUMDAR_4}
{S.\,N.~Majumdar}, {A.\,J.~Bray}, {S.\,J.~Corwell}, and C. Sire,
  Phys.~Rev.~Lett. {\bf 77},  3704  (1996).

\bibitem{CUEILLE_1}
S. Cueille and C. Sire, J.~Phys.~A {\bf 30},  L791  (1997).

\bibitem{MAJUMDAR_5}
{S.\,N.~Majumdar}, Curr.~Sci.~India {\bf 77},  370  (1999).

\bibitem{DERRIDA_2}
B. Derrida, V. Hakim, and V. Pasquier, Phys.~Rev.~Lett. {\bf 75},  751  (1995).

\bibitem{LEE_1}
{B.\,P.~Lee} and {A.\,D.~Rutenberg}, Phys.~Rev.~Lett. {\bf 79},  4842  (1997).

\bibitem{OERDING_1}
K. Oerding and {F.~van\,Wijland}, J.~Phys.~A {\bf 31},  7011  (1998).

\bibitem{HINRICHSEN_2}
H. Hinrichsen and M. Antoni, Phys.~Rev.~E {\bf 57},  2650  (1998).

\bibitem{ALBANO_1}
{E.\,V.~Albano} and {M.\,A.~Mu\~{n}oz}, Phys.~Rev.~E {\bf 63},  031104  (2001).

\bibitem{ZIFF_1}
{R.\,M.~Ziff}, E. Gulari, and Y. Barshad, Phys. Rev. Lett. {\bf 56},  2553
  (1986).

\bibitem{ROSSI_1}
M. Rossi, R. Pastor-Satorras, and A. Vespignani, Phys.\,Rev.\,Lett. {\bf 85},
  1803  (2000).

\bibitem{LUEB_19}
S. L{\protect\"u}beck, Phys.~Rev.~E {\bf 64},  016123  (2001).

\bibitem{HINRICHSEN_1}
H. Hinrichsen, Adv.~Phys. {\bf 49},  815  (2000).

\bibitem{HINRICHSEN_3}
H. Hinrichsen and {H.\,M.~Koduvely}, Eur.~Phys.~J.~B {\bf 5},  257  (1998).

\bibitem{LUEB_20}
S. L{\protect\"u}beck and A. Hucht, J.~Phys.~A {\bf 34},  L577  (2001).

\bibitem{GRIMMETT_1}
{G.\,R.~Grimmet} and {D.\,R.~Stirzaker}, {\em Probability and Random Processes}
  (Clarendon Press, Oxford, 1992).

\bibitem{JENSEN_3}
I. Jensen and R. Dickman, Phys.~Rev.~E {\bf 48},  1710  (1993).

\end{thebibliography}

\begin{table}[ht]
\caption{The critical exponents 
of the CLG model for various dimensions~$D$.
The symbol $^{\ast}$ denotes logarithmic corrections
to the power-law behavior.
The values for $\rho_{\rm c}$, $\beta$, $\gamma'$ 
are obtained from~\protect\cite{LUEB_19,LUEB_20}.
The values of $\nu_{\scriptscriptstyle \perp}$
are obtained from the scaling relation
$\gamma'=\nu_{\scriptscriptstyle \perp} D - 2 \beta$~\protect\cite{JENSEN_3}.
}
\label{table:exponents}
\begin{tabular}{lllllllll}
$D$       &  $\rho_{\rm c}$	& $\beta$	& $\gamma'$ 
& $\nu_{\scriptscriptstyle \perp}$  & $\nu_{\scriptscriptstyle \parallel}$
& $\nu_{\rm l}$ & $\theta_{\rm g}$ & $\theta_{\rm l}$\\  
\hline \\
%
%
$2$     &  $0.34494$	& $0.637$	& $0.384$  
& $0.83$ & $1.21$ & $1.10$ & $1.58$ & $1.41$\\ 
%
%
$3$     &  $0.2179$	& $0.837$ & $0.18$  
& $0.62$ & $1.04$ & $1.04$ & $1.54$ & $1.19$\\
%
%
$4$     &  $0.1571$	& $1^{\ast}$      & $0^{\ast}$ 
& $1/2$ & $1$ & $1$ & $1.5$ & $1$\\
%
%
$5$ 	&  $0.1230$	& $1$             & $0$ 
& $1/2$ & $1$ & $1.075$ & $1.5$ & $1$\\     
\end{tabular}
\end{table}

\end{document}